\documentclass[10pt]{article}
\usepackage{graphicx, verbatim, url, amssymb, amsmath, amsfonts, amsthm, latexsym, wrapfig}

\begin{document}

\title{Crypto art: A decentralized view}

\author{
Massimo Franceschet \\
University of Udine \\ 
\url{massimo.franceschet@uniud.it} \and 
Giovanni Colavizza \\
University of Amsterdam and The Alan Turing Institute \\ 
\url{g.colavizza@uva.nl} \and 
T'ai Smith and Blake Finucane \\
Department of Art History, Visual Art and Theory \\
University of British Columbia \\
\url {tai.smith@ubc.ca}  \\ 
\url{blake@blakefinucane.ca} \and 
Martin Lukas Ostachowski \\
Art by MLO \\
\url{martin@ostachowski.com} \and \newline
Sergio Scalet \\
Hackatao \\
\url{sergio.scalet@hackatao.com} \and
Jonathan Perkins \\
SuperRare.co Co-founder \\
\url{jperkins@pixura.io} \and 
James Morgan\\
KnownOrigin.io Co-founder \\
\url{james.morgan@blockrocket.tech} \and
Sebasti\'{a}n Hern\'{a}ndez \\
The Momus Collective \\
\url{sebdecentr@protonmail.com}
}
\maketitle

\section*{Prologue}

This is a \textit{``decentralized position paper"} on crypto art, which includes viewpoints from different actors of the system: artists, collectors, galleries, art scholars, data scientists. The writing process went as follows: a general definition of the topic was put forward by two of the authors (Franceschet and Colavizza), and used as reference to ask to a set of diverse authors to contribute with their viewpoints asynchronously and independently. No guidelines were offered before the first draft, if not to reach a minimum of words to justify a separate section/contribution. Afterwards, all authors read and commented on each other's work and minimal editing was done. Every author was asked to suggest open questions and future perspectives on the topic of crypto art from their vantage point, while keeping full control of their own sections at all times. While this process does not necessarily guarantee the uniformity expected from, say, a research article, it allows for multiple voices to emerge and provide for a contribution on a common topic. The ending section offers an attempt to pull all these threads together into a perspective on the future of crypto art. We briefly introduce the contributors in what follows.

\textit{Massimo Franceschet} is an experienced data scientist. He teaches Data Science at the Bachelor's degree in Internet of Things, Big Data and Web and Network Science at the Master's degree in Computer Science at the University of Udine in Italy. As a researcher, he has published 60 peer-reviewed publications in the fields of data science, complex networks, bibliometrics, logic and artificial intelligence. Under the name \textit{HEX0x6C}, he is also a generative artist, with works exhibited on different galleries including SuperRare and KnownOrigin.  

\textit{Giovanni Colavizza} is an assistant professor of digital humanities at the University of Amsterdam. He was previously part of the research engineering group of The Alan Turing Institute and of the quantitative science studies group at the Centre for Science and Technology Studies (CWTS), Leiden University. Giovanni works on machine learning and data science applied to GLAM collections (Galleries, Archives, Libraries and Museums) and on the use of computational methods in the humanities.

\textit{T'ai Smith} is an associate professor of art history and media theory at University of British Columbia in Vancouver, Canada. Her research focuses on media and economic concepts in art and design. Author of \textit{Bauhaus Weaving Theory} (University of Minnesota Press, 2014), she has published over 30 articles, chapters, and museum catalogue essays, including for the Museum of Modern Art and Tate Modern. She is currently at work on two book manuscripts: \textit{Fashion After Capital} and \textit{Textile Media: Tangents from Contemporary Art and Theory}. 

\textit{Blake Finucane} graduated with her master’s degree in Art History and Theory from the University of British Columbia, where she produced one of the first academic theses on crypto art and blockchain technology in 2018, under the guidance of her supervisor professor T'ai Smith. She is now working at a private investment firm that specializes in financing early stage companies.

\textit{Martin Lukas Ostachowski} is a geometric abstraction artist based in Quebec, Canada. He explores blockchain technology through the subject of clouds in physical and digital mediums. His upcoming solo exhibition entitled \textit{Tropopause Contemplation -- Blockchain Technology and Inclusive Decentralization} in September 2019 is one of the first blockchain and crypto art-themed solo exhibitions in North America.

\textit{Sergio Scalet}. Hackatao merges the cultured quotations from the past to bold and ultracontemporary forms, standing out as an innovative exponent of the current artistic scenario like crypto art. The artistic diptych Hackatao is formed by Sergio Scalet and Nadia Squarci. The duo was formed in Milan in 2007 working together with the creation of the Podmork, sculptures with soft and totemic forms, which are at the center of their imaginative research. Hackatao has participated in numerous solo and group exhibitions in Italy and around the world. They realized the first exhibition on crypto art in December 2018, at \textit{Palazzo Frisacco} in \textit{Tolmezzo} (Italy), involving several artists part of the movement.

\textit{Jonathan Perkins} is an American entrepreneur whose work has spanned technology and art, as a performing musician, recording engineer, computer programmer and creative technologist. In 2018, he co-founded SuperRare, a social platform for artists and collectors to trade crypto art. He holds a bachelor’s degree in Digital Media Arts from San Francisco State University.

\textit{James Morgan}. Having been involved in the digital arts space for only 18 months I feel I still have lots of learn but recognise how far we have come as a team and myself personally. The tech, the people and artwork has shown me the power of what a fairer system could look like and I hope to help push this forward in the future. It's been a pleasure to work on KnownOrigin and I hope for its success in the future.

\textit{Sebasti\'{a}n Hern\'{a}ndez} is a technologist and entrepreneur who happens to love physical art. His involvement in The Momus Collective -- a group of crypto enthusiasts who build Virtual Reality experiences -- led him to purchase over 300 crypto art tokens through multiple providers including SuperRare, KnownOrigin, snark.art, CryptoArte.io and Larva Labs.

\section{Crypto art: Rare digital art on the blockchain}

Rare digital art, also known as \textit{crypto art}, is limited-edition collectible art cryptographically registered with a token on a blockchain. Tokens represent transparent, auditable origin and provenance for a piece of digital art. Blockchain technology allows tokens to be held and securely traded from one collector to another.

Digital galleries like SuperRare \cite{superrare} and KnownOrigin \cite{knownorigin} exhibit rare digital artworks, such as static and animated images.  When a digital asset made by an artist is added to a digital gallery, a token is generated by a smart contract and deposited in the artist's wallet. The token is permanently linked to the artwork, and is a unique, one-of-a-kind asset that represents ownership and authenticity of the underlying artwork. 

Once created, the artwork starts its life on the given blockchain, where a fan or collector can purchase it, and where it can be subsequently exchanged, traded or held by collectors like any other rare artifact. Typically, artworks can be sold using auctions: creators list a piece of art for sale, bidders make offers, and the creator has the ability to accept the bid or not. When an asset is sold, it is directly transferred to the buyer's wallet, while the corresponding price in crypto currency is moved to the seller's wallet. Thanks to the blockchain, each transaction is cryptographically secured and peer-to-peer, meaning neither the funds nor the asset are ever held by the gallery or any other third party.

Crypto art rests on the shoulders of blockchains. In the following, we briefly review this new-old technology. Then, we give a real example of the work flow of a crypto art work.

\subsection{Blockchain}

A blockchain is a distributed system using cryptography to secure an evolving consensus about an economically valuable token \cite{NIST18}. Notably, the idea is not new and dates back to 1991 when Haber and Stornetta proposed this technology to certify the date of the documents and avoid tampering \cite{HS91}. In the abstract of their farsighted paper, the authors write:

\begin{quote}
\textit{The prospect of a world in which all text, audio, picture, and video documents are in
digital form on easily modifiable media raises the issue of how to certify when a document was created or last changed. The problem is to time-stamp the data, not the medium.}
\end{quote} 

The idea went mostly unnoticed until it was adopted by Satoshi Nakamoto in 2008 to create a (now popular) crypto-currency known as Bitcoin \cite{Nakamoto08}. The conclusion of Nakamoto's white paper is intriguing:

\begin{quote}
\textit{We have proposed a system for electronic transactions without relying on trust. [...] we
proposed a peer-to-peer network using proof-of-work to record a public history of transactions
that quickly becomes computationally impractical for an attacker to change if honest nodes
control a majority of CPU power. The network is robust in its unstructured simplicity.}
\end{quote}

More specifically, a blockchain is a chain of blocks, where a block is a container for data. In its simplest form a block contains: 

\begin{enumerate}
\item a hash of the current block, that identifies the block,
\item a timestamp of block creation, 
\item a list of transactions, and 
\item the hash of the previous (parent) block in the chain. 
\end{enumerate}

Hashes of blocks are created using a \textit{cryptographic hash function}, that is a mathematical algorithm that maps data of arbitrary size to a bit string of a fixed size (called hash or digest). The hash is used to certify the information content of the block: by modifying even a single bit of the content, the hash completely changes. Furthermore, the digest of the current block is computed in terms of the previous block digest, which means that changing one block invalidates all future blocks. Hence if you alter one block you need to modify not only its hash, but also that of all following blocks for the chain to be valid. 

Hashes alone are not enough to prevent tampering, since hash values can be computed fast by computers. A \textit{proof-of-work} algorithm controls the difficulty of creating a new block. Blocks are created (or mined) by so-called miners. When a new block has to be created, a hard computational problem, that can be attacked only with a brute-force approach, is sent out to all potential miners. The first miner that solves the problem creates the block and gets a reward. Hence, to modify a past block, an attacker would have to redo the proof-of-work of the block and all blocks after it and then catch up with and surpass the work of the honest nodes. However, the probability of a slower attacker catching up diminishes exponentially as subsequent blocks are added \cite{Nakamoto08}.

Typically, the data field of a block contains a certain number of \textit{transactions}. Each transaction has a sender, a receiver, a token that is transferred from sender to receiver, and a fee of the transaction. A \textit{token} is a quantified unit of value recorded on the blockchain (such as a coin or an artwork). Transactions are stored in a pool of pending transactions. Each mined block will include the most profitable transactions (the ones with the highest fees) among the pending ones. The miner of the block gets as reward the sum of the fees of transactions in the block plus additional cryptocurrency as a compensation for the  CPU time and energy consumed to perform the proof-of-work. Furthermore, blockchain uses asymmetric cryptography to implement \textit{digital signatures} of transactions. Each transaction is signed by the sender with their private key and anyone can verify the authenticity of the signature (and of the transaction) using the sender's public key.

The incentive of rewards may help encourage nodes to stay honest. As Nakamoto claims in the white paper \cite{Nakamoto08}: 

\begin{quote}\textit{If a greedy attacker is able to assemble more CPU power than all the honest nodes, he would have to choose between using it to defraud people by stealing back his payments, or using it to generate new coins. He ought to find it more profitable to play by the rules, such rules that favour him with more new coins than everyone else combined, than to undermine the system and the validity of his own wealth}.
\end{quote}

Finally, the blockchain ledger is distributed over a \textit{peer-to-peer network}. The steps to run the network are as follows \cite{Nakamoto08}:

\begin{enumerate}
\item new transactions are broadcast to all nodes;
\item each node collects new transactions into a block;
\item each node works on finding a difficult proof-of-work for its block;
\item when a node finds a proof-of-work, it broadcasts its block to all nodes;
\item nodes accept the block only if all of its transactions are valid and not already spent;
\item nodes express their acceptance of the block by working on creating the next block in the
chain, using the hash of the accepted block as the previous hash.
\end{enumerate}

In conclusion, the blockchain brings together mathematics (cryptography), computer science (distributed systems), economics (exchange of tokens with economic value) and politics (mechanisms for reaching consensus).  Blockchain technology aims to replace trust we (usually) place in intermediaries through a smart system that combines the objectivity of mathematics, the neutrality of algorithms and the economic incentive of humans to play by the rules. It is much more than a technology, it is also a culture and community that is passionate about creating a more equitable world through decentralization. As Bettina Warburg states in her TED talk\footnote{\url{https://www.ted.com/talks/bettina_warburg_how_the_blockchain_will_radically_transform_the_economy}.}: 

\begin{quote}\textit{We are now entering a radical evolution of how we interact and trade because, for the first time, we can lower uncertainty not just with political and economic institutions but with technology alone}. 
\end{quote}

\subsection{How crypto art works}

\begin{figure}[t]
\begin{center}
\includegraphics[scale=0.10, angle=0]{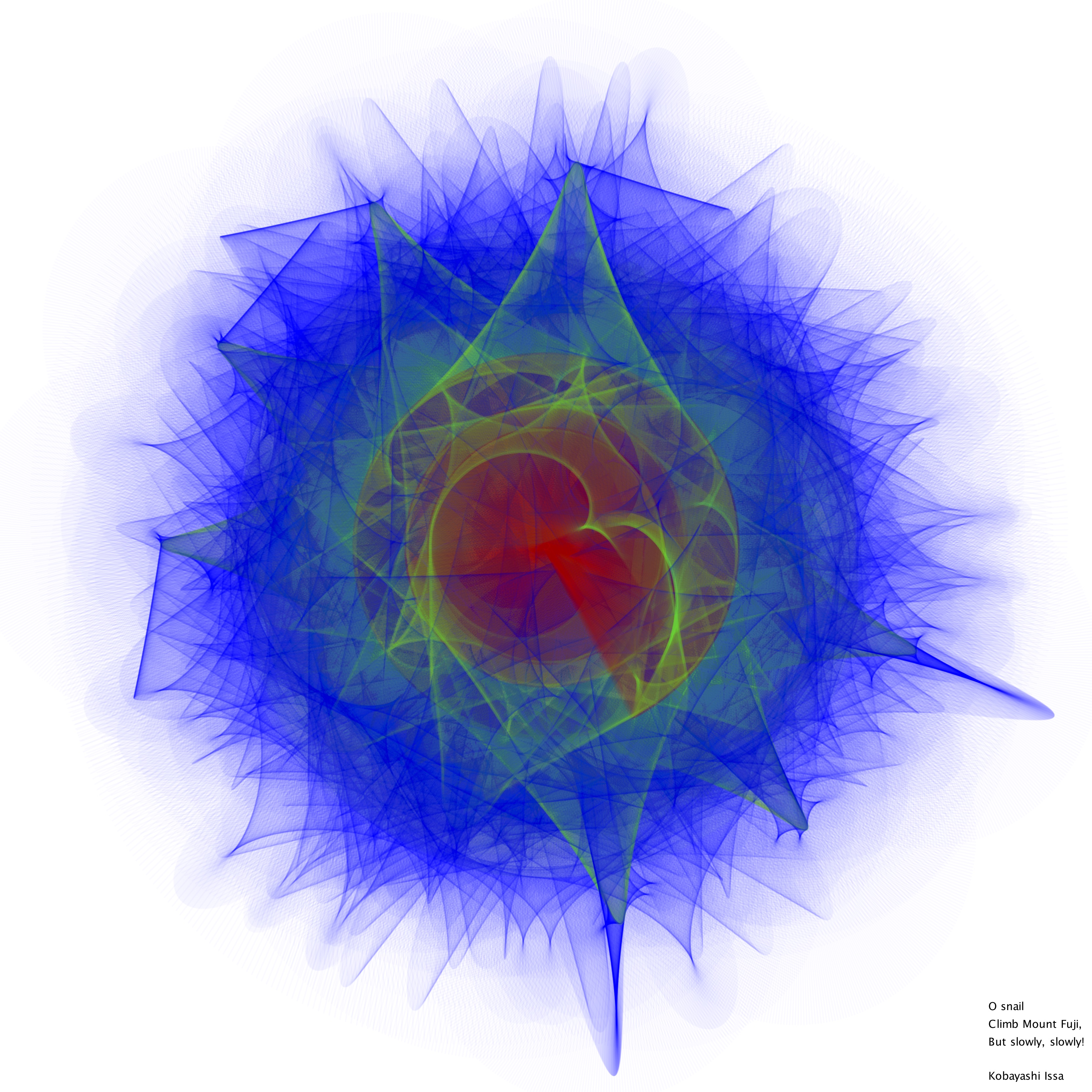}
\end{center}
\caption{O Snail, from crypto artist HEX0x6C, is exhibited in the SuperRare gallery.}
\label{fig:snail}
\end{figure}

Crypto art is a very recent artistic movement in which the artist produces works of art, typically still or animated images and often in close collaboration with the machine (not necessarily a computer but also, for example, a scanner or an old Polaroid) and distributes them using blockchain technology and the peer-to-peer InterPlanetary File System (IPFS) network.

Sergio Scalet, one of the first-movers into crypto art territory and co-author here, speaks about this trend as follows:\footnote{\url{https://www.coiners.it/cryptoart-il-nuovo-paradigma-dell-arte-ultracontemporanea} (in Italian).}

\begin{quote}
\textit{Crypto art is one of the few artistic movements of this millennium that speaks the language of its time;  it allows new digital expression languages to have a fast, pulsing, tailor made market on the new generation of collectors. If until some time ago an artist who created digital artworks had very little chance to put them on the market, for the obvious reason of their infinite reproducibility, today thanks to blockchain technology these digital artworks are created in a certified small number of editions and hence acquire an exchangeable value. Crypto artworks remain infinitely reproducible and visible to all, but only the purchasing collector owns what the artist will call the original, the unique token of the work. Obviously, the collector must accept the perspective of owning a digital work and not a physical object to hang on the wall. A concept much more understandable to the Millenials than to the generation of Boomers still linked to the material, weighty palpability of a work of art.}
\end{quote}

But how does crypto art really work? We'll make a real example with respect to the SuperRare gallery. Consider the artwork entitled \textit{O Snail} \cite{Snail}, shown in Figure \ref{fig:snail}. The artwork, exhibited in the crypto art gallery SuperRare \cite{superrare}, represents an ancient haiku of Kobayashi Issa. 

When the artist uploads the artwork in the SuperRare gallery, a transaction is created in the blockchain Ethereum. The transaction creates and transfers a token uniquely associated with the work of art into the artist's cryptographic portfolio. The transaction is digitally signed by the artist, using asymmetric encryption, in order to prove the authenticity of the work.

The gallery distributes the artwork (that is the JPEG image in this case) over the nodes of the IPFS peer-to-peer network \cite{ipfs}. The IPFS network names the image with a unique code that uniquely matches its content. This means that the same image, even if distributed over several nodes of the network, will always have the same name and will be conceptually identified as a single resource (unlike the usual Web we know). The IPFS website states:\footnote{\url{https://ipfs.io}.}

\begin{quote}
\textit{IPFS and the Blockchain are a perfect match! You can address large amounts of data with IPFS, and place the immutable, permanent IPFS links into a blockchain transaction. This timestamps and secures your content, without having to put the data itself on the chain.}
\end{quote}

The digital work now begins its immortal life on the blockchain. Anyone can review it and possibly buy it at the price set by the artist (many sales work by auction). The price is expressed in a crypto currency called Ether, one of the major digital currencies by market capitalisation. At the time of purchase, a new transaction is placed in Ethereum blockchain: the token of the artwork passes in the portfolio of the buying collector while the agreed amount in Ether moves in the portfolio of the selling artist. The artwork, even after being sold, remains on the market and can still be traded. In some marketplaces, each secondary market sale rewards the original artist as a percentage (10\% of the sale price on SuperRare). All these steps, which in the traditional art market occur in months or years, thanks to blockchain technology can take place in the matter of moments, in a certified and secure way.


As Jason Bailey states in his introduction to the first Italian exhibition of crypto art (Tolmezzo, December 2018 - February 2019. Full text available upon request):

\begin{quote}
\textit{Unlike the traditional art world, (crypto) artists did not seek permission from gallerists, agents, auction houses, or other gatekeepers to share and sell their work. Instead, leveraging the blockchain, they simply decided on their own to show their work and make it available. Naturally, the old guard regularly speaks derisively of crypto art, but if history is any measure, this is a sure-fire signal that crypto artists are on the right path.}
\end{quote}

\section{Viewpoints}

The authors of these subsections are as follows:
\begin{itemize}
    \item \textit{The data scientists}: Massimo Franceschet and Giovanni Colavizza.
    \item \textit{The art scholars}: T'ai Smith and Blake Finucane.
    \item \textit{The artists}: Martin Lukas Ostachowski and Sergio Scalet.
    \item \textit{The gallerists}: SuperRare (Jonathan Perkins) and KnownOrigin (James Morgan).
    \item \textit{The collectors}: Sebasti\'{a}n Hern\'{a}ndez.
\end{itemize}

\subsection{The data scientists}

\textit{Big data} is typically characterized by the three Vs of volume, velocity, and variety \cite{3V}:

\begin{itemize}
\item \textit{Volume} of data refers to the size of data managed by the system. 
\item \textit{Velocity} is the speed at which data is created, accumulated, ingested, and processed.
\item \textit{Variety} refers to the fact that big data includes structured, semi-structured, and unstructured data. Structured data can be represented using a data model, such as the relational model, in which data are organized in the form of tables containing rows and columns. Unstructured data have no identifiable formal structure, and include text, audio, video, images. Some forms of unstructured data may fit into a format that allows well-defined tags that separate semantic elements; this format may include the capability to enforce hierarchies within the data. This kind of data is referred to as semi-structured data.
\end{itemize}

Crypto art possesses at least two of the Vs of big data: velocity and variety. Sequences of crypto art events, such as creations, bids, and sales, take place at the fine time granularities of minutes and even seconds. This defines crypto art as a \textit{real-time stream} of events, more similar to financial trading than traditional art auctioning.  Moreover, data in crypto art is made of:

\begin{itemize}
\item \textit{artworks}, typically static or animated images (unstructured data);
\item \textit{metadata} about the artworks, such as author, title, description, keywords (semi-structured data);
\item \textit{events}, such as valued bids and purchases with the relative timestamps (structured data). 
\end{itemize}

We stress that all these data are fully openly available: this is a crucial advantage of crypto art if compared to traditional art, from our point of view. As (data) scientists, we consider the following questions in order to start shaping a ``crypto art analytics'' research area:

\begin{itemize}
\item What are the similarities and discrepancies of crypto art with respect to the scholarly publication system? In particular, can we adapt and apply methods from the quantitative analysis of science (scientometrics) to the analysis of crypto art? By casting a comparison with another well-known system, we aim to elucidate the characteristics of interest of crypto art.
\item What are interesting problems regarding crypto art or which can be considered using crypto art data? Do these problems call for new methods? Using the starting point of the comparison with the scholarly publication system, we aim to broaden the set of questions of interest, as well as suggest directions for related methodological contributions. 
\item What is the state of the related literature on art analytics, more generally?
\end{itemize}

\subsubsection{A comparison with the scholarly publication system}

Let us compare the work flow of a scholar with that of an crypto artist, by distinguishing three sequential phases: creation, presentation and endorsement.

\begin{enumerate}
\item \textit{Creation}. The artist creates an artwork while the scholar writes a scientific contribution, such as a journal article. Both outcomes, the artwork and the article, are works of creativity and need to be original with respect to the state-of-the-art. Both the artist and the scholar start from an idea and then develop it into a piece of art or scholarship. Typically, the artist works alone while the scholar collaborates with peers, even though there are artists who create within collectives and scholars that prefer to work individually. Oftentimes and increasingly, both the artist and the scholar interact with a computer, the artist for generating the artwork (for instance using randomness, complexity or artificial intelligence) and the scientist for making experiments or analysing data. 

\item \textit{Presentation}. When the work is ready, the artist uploads and exhibits it in an online gallery, while the scholar submits it to a conference or journal. Here we have a divergence. The scholar's work is reviewed by peers and accepted for publication only if the article is found to be sound and sufficiently original with respect to the state-of-the-art. All scholars can submit an article to any journal: it is the work that will be reviewed, not its author(s). On the other hand, a crypto gallery reviews the artist, not the artwork. If artists meet the gallery's standards, they are white-listed once and for all by the gallery and can upload any artwork without further review (as long as these works do not violate the terms of service of the gallery). When the work of art is exhibited, anyone can view and download it. On the other hand, if a scientific work is accepted for publication, it is visible only to subscribers of the journal or attendees of the conference, although open access is increasingly being adopted. Scholars can submit different works to different journals; similarly artists can exhibit their art in different galleries.  

\item \textit{Endorsement}. Both works of art and scientific contributions can be endorsed by the respective communities, gaining in popularity. A scientific article is endorsed, for example, when a peer references it in another article. The number of citations from other scholars accrued by an article is an indicator of its popularity within the community. An artwork is endorsed when a collector makes a valued bid (during an auction sale) or a collector directly purchases the piece for a given price. Similarly, the number of bids made for the artwork, or the number of times the artwork is traded among collectors, might be considered as an indicator of the popularity of the piece of art. Interestingly, all endorsements for an artwork are valued (they have an amount of crypto currency associated with them), and this value might be considered as an estimate of the value of the artwork at a given time. On the other hand, citations are typically not weighted (they do not come with an associated value). In both science and art there exist self-endorsements: scholars citing their own articles and artists buying (maybe with different identities) their own artworks. In both cases, these self-references do not contribute to the impact of the individual. Furthermore, besides popularity, one can also investigate the prestige of the works of art and of scholarly publications. We might argue that a bid to an artwork made by a prestigious collector, or a citation to an article given by an authoritative scientist, are more important than endorsements given by anonymous individuals. This parallelism between the endorsement system in art and science is quite precious since the scholarly publication system has experience in methods and tools to quantitatively assess the popularity and prestige of a scientific contribution using citations.
\end{enumerate} 

Crypto art analytics would focus first and foremost on the endorsement stage, as this is where data are more abundant and a stronger parallel with the scholarly publishing system can be cast. The creative process is intrinsically difficult to observe and study, while the presentation stage offers crucial distinctions which ought to be considered and can be analysed via comparisons. To summarize, the following parallels can be cast:

\begin{itemize}
	\item \textit{Scholarly contributions and works of art}: these are the tangible objects encapsulating creative work.
	
	\item \textit{Authors and artists}: the creators of works of creativity. Creators gain reputation out of the recognition given to their work, and transmit their accumulated reputation to future works.
	
	\item \textit{Journals and galleries}: where works of creativity are gathered and put in display. Both journals and galleries also gain from and contribute to the reputation of creators and their works.
	
	\item \textit{Citations and bids}: with the important difference that bids are weighted by the (crypto currency) value being offered, citations are not.
	
	\item \textit{Altmetrics}: both systems possess a variety of alternative metrics, such as view counts, downloads, shares, likes that might complement the picture of impact of the scholarly or artistic work.
\end{itemize}

A general difference between crypto art and the scholarly publication system, one that involves all phases of creation, presentation and endorsement is \textit{velocity}. For a scientist, it takes considerable time (typically months or even years) to conceive and write an article, submit it to a journal, wait for peer reviews, modify the article according to the comments of the reviewers, re-submit the article and possibly wait for additional comments from reviewers, and eventually publish the final version in the journal. Moreover, citations to the article typically arrive years after publication. On the contrary, the workflow of crypto art is potentially very fast: having the right idea and using a generative computer-aided process, the author can quickly produce one piece, instantaneously upload and exhibit it (if the author is white-listed for the gallery), and, if lucky, sell it in a matter of minutes. After the sale, the artwork can be traded in the secondary market (even outside the gallery) with the same speed. We might say that the working time granularity in science is years, while the time granularity in crypto art is much finer. 

The parallel between crypto art and science breaks down on other important aspects. 
In the art world, \textit{collectors} (or investors) can place bids and exchange works of art irrespective of them also being artists. Conversely, only scholarly peers can give citations by producing new works of scholarship. The role of collectors is crucial, as they provide for an external, non-peer source of recognition to artists, who can then be considered to primarily look for high-bid collectors instead of peer recognition. Finally, artworks can be gifted or exchanged for other artworks, without bids and purchases being involved. They can even be used as currency themselves (in fact, crypto artworks and crypto coins are made of the same matter).

\subsubsection{Problems in crypto art analytics}

We propose three main problem areas in crypto art analytics:
	
\textit{Rating and ranking}. By taking advantage of the similarities between the endorsement system in science and crypto art, we can borrow and adapt metrics from the quantitative analysis of science. The general goal here is to provide for a ranking of actors according to given criteria, for example of the leading artists by value of works sold over a certain time. In general, metrics should be informative for galleries, art investors, as well as the artists themselves. General questions which can be answered via well-defined crypto art metrics are, for instance, what are the trending artists to showcase for a gallery? What are the emerging artists to invest in for an investor? What are the top collectors to get in touch with for an artist? An interesting challenge here is that the crypto art sales network is not static but evolves over time, adding new nodes (actors) and links (sales) every hour. Moreover, sales are typically valued in a given crypto currency, whose counterpart in fiat currency might undergo large oscillations. While the use of stablecoins (DAI, USDC) is growing, this might call for metrics that consider event timing. It is worth stressing that these crypto art metrics would measure commercial success and popularity, which are informative but do not equate to artistic quality, in the same way as citations to an article are an indicator of its impact and not necessarily a measure of its overall merit. 

\textit{System analysis and modeling}. The whole crypto art market can be put under analysis. The goal is to understand it as a system, instead of just over the specific parts captured by metrics. Examples of interesting questions include: what are the different types of actors (e.g. artists, collectors, investors, by-standers) and how do they interact? Is the market expanding or declining, and if so in which directions? Can the primary or secondary market be considered mature or not? More generally, the ultimate goal of this problem area would be to formally model the art system, capturing some desired aspects and the interactions among actors in it. The main fields from where to borrow techniques in this case would be network science and complex systems. Some properties of the art market have been considered in previous literature. Liu et al.\ \cite{LWSGSW18} explore hot streaks in artistic (artists), cultural (film directors) and scientific careers. For artists, they use a dataset collected from the Artprice and Findartinfo services.\footnote{\url{https://lu-liu.github.io/hotstreaks}.} It contains 3480 artists with at least 15 works listed and a career of at least 10 years. The main finding is that a simple transition model (in and out the hot streak) explains the data well. Most careers show one, sometimes two hot streaks, which can happen at any time during it.

\textit{Prediction of success}. Given a model of the system, it is possible to forecast its evolution in general, or for a specific aspect of interest, for example by providing a list of emerging artists by future value of their artworks. This problem area focuses on predicting the evolution of the crypto art system and, in particular, predict success within it. Several previous works have considered the problem of success prediction in art. For example, Fraiberger et al.\ \cite{FSRRB18} use data from Magnus, the self-labeled Shazam for Art.\footnote{\url{http://www.magnus.net}.} These data, among other information, include 127,208 auctions (although this specific dataset appears not to be used in the article). The main finding is that there are strong lock-in mechanisms that drive artistic retention and reputation, which are linked to being able to exhibit in the network of high-prestige galleries. Mitali and Ingram \cite{MI18} build a social structural model of fame, which departs from the atomistic view of prior literature where creativity is the sole driver of fame in creative markets. Their dataset consists of 90 pioneers of the early 20th century abstract art movement (1910-1925). Their main finding is that artists with a large and diverse network of contacts were most likely to be famous, regardless of how creative their art was. Art creativity was measured using both objective computational methods (machine learning techniques) and subjective expert evaluations. Sigaki et al.\ \cite{SPR18} present a quantitative analysis of almost 140,000 paintings, spanning nearly a millennium of art history. Based on the local spatial patterns in the images of these paintings, they estimate the permutation entropy and the statistical complexity of each painting. These measures map the degree of visual order of artworks into a scale of order -- disorder and simplicity -- complexity that locally reflects qualitative categories proposed by art historians.


In summary, the velocity and variety of data in crypto art, as well as their full availability, open up compelling avenue for future scientific work. The very fact that the crypto art market blends auction houses with financial trading, makes it a novel phenomenon, all the same representative of creative industries in a fully digital space. 

\subsection{The art scholars}

\subsubsection{Historical perspective}

Art is often conflated with the idea of \textit{beauty}, but this is an imprecise concept, at best. It must be recognized that the achievement of beauty as an aesthetic category is historically specific -- subject to conventions and fashions; and further, that the achievement of beauty is not always the artist's goal. Alternative objectives can include the edification or enlightenment of the viewer on a topic or question about society, politics, or the discourse of art itself -- for example, intellectual debates about abstraction, its systems of representation, or the institutional determinants of economic or qualitative value.

Indeed, such interrogations of value (whether monetary or moral) have been the focus of scrutiny by many artists since the early twentieth century. Oliver Roeder cites Andy Warhol as the most notable artist whose work acted in similar ways to crypto art. This is an obvious choice, as Roeder explains \cite{Roeder}: \begin{quote}\textit{No one artist's work was more responsible for melding art with cash than Andy Warhol’s, which generates great mass media appeal and exists in great quantities.}\end{quote} Warhol openly proclaimed his affinity for commercial success and making money, looking to promote his art and brand to the widest audience possible. Crypto art does of course evoke this model, melding currency with art creation, and there is a lot that could be written about the relationship between pop art and crypto art. Similarly, one might look to the history of digital art (or ``software art" since the 1960s, or ``Internet art" since the 1990s) to understand this new type of practice. 

We have chosen, however, to focus on a genealogy that stems from conceptual art and dadaist Marcel Duchamp, especially those works that sought to foreground the commercial tendencies of the art world by actively engaging (or exploiting) its institutional frameworks or codes, and the legal contract in particular. Many characteristics that were endemic to Duchamp's practice, and later conceptual art in the 1960s and seventies, are now visible in the immaterial and distributive logic of crypto art, though the discourse around the latter has not yet focused on this history. 

\subsubsection{A genealogy: Conceptual art and Marcel Duchamp}\label{sec:geneaology}

The \textit{dematerialization of art}, a phrase that was coined in 1967 by critics Lucy Lippard and John Chandler and elaborated in their 1968 article of the same name for \textit{Art International} magazine, identified the quintessential features of conceptual practices \cite{Lippard}. As Lippard later defined it, conceptual art was, essentially \cite[vii]{Lippard2}: \begin{quote}\textit{work in which the idea is paramount and the material form is secondary, lightweight, ephemeral, cheap, unpretentious and/or ``dematerialized".}\end{quote} It was able to be created and viewed outside of major art centers (like New York or Paris) or museums and art galleries, features which helped conceptual artists challenge the elitist tendencies of the art world as well as how art could be bought and sold. Lippard explains \cite[xviii]{Lippard2}: \begin{quote}\textit{the easily portable, easily communicated forms of conceptual art made it possible for artists working out of the major art centers to participate. [..] They could also carry their work with them as they moved around the country or world.}\end{quote} Blockchain technology permits a different kind of immaterial object, but many of the same capabilities apply. The digital and contractual determination of crypto art works in similar ways: decentralizing art creation and sales allows for the hyper-portability of the ``object" and the rejection of conventional markets and institutions.

In the early sixties, conceptual artists began to embrace legal and administrative language as a substitute for traditional visual content. These artists were captivated by bureaucratic forms and procedures, and the contract became an effective instrument to explore and express new definitions of art. It was also a means of drawing attention to the difficulties of being an artist in a market system and the demands in interfacing with the business side of the industry. The art duo N.E. Thing, Co. from Vancouver, Canada, for instance, legally registered as a corporation in order to focus, self-reflexively, on the administrative nature of their artistic practice. Meanwhile, the New York-based Joseph Kosuth contended that ``Aesthetics [..] are conceptually irrelevant to art."\cite{Kosuth} He did not think that something needed to be aesthetic (by which he meant sensually material) for it to be considered art; it should, in his mind, simply express ``Art as Idea as Idea." The intention was to challenge an art object's visuality, commodity status, and/or its form of distribution. By using cheap and easily distributed media -- e.g., bound books with photographs and text, ``event scores" written on cards, contracts, and other forms of printed matter -- conceptual artists attempted to discharge any association with transcendence or the idea that the work is a precious and unique product of a skilled hand. In certain respects, this is similar to crypto art, as the visual aspects of the image are not especially complex and can be grasped without pondering the artist's use, say, of colour and form. Some crypto artists might even contend their work is more about the idea -- the fact that it can be transmitted virtually -- than what the image itself represents. Importantly, however, crypto art essentially emulates, through the mechanism of encryption, the modernist (or early twentieth-century) fixation on the unique and original work of art \cite{Krauss}.

In his 1990 article ``Conceptual Art 1962-1969: From the Aesthetic of Administration to the Critique of Institutions," art historian Benjamin H.D. Buchloh posits that ``the introduction of legalistic language and an administrative style of the material presentation of the artistic object" in conceptual art in fact began earlier in the twentieth century, with Marcel Duchamp in 1944 \cite[118]{Buchloh}. This was the year that Duchamp hired a notary to inscribe a statement of authenticity on his ``rectified readymade" \textit{L.H.O.O.Q.} (1919) -- essentially a poster print of Leonardo's Mona Lisa, with an added moustache and goatee. The notary signed the work more than two decades after it was initially produced, affirming that it was the ``original `ready-made.'" This was a practical move, validating the image's authenticity, which could (ostensibly) increase its sale price. But this act \textit{also} challenged the definition of a ``work of art," since the object's value was inscribed via a bureaucratic procedure, rather than through a connoisseur's analysis, say, of the artist's ``hand" or style of mark making with a brush or pen. Nevertheless, this example illuminates the central role that both authenticity and provenance (the history of ownership) play in the notion of artistic value in art and how important these factors have been historically. 

Indeed, artists have been explicitly examining this relationship between value, authenticity and the definition of art for about a century. Duchamp's \textit{Tzanck Check} (1919) and \textit{Monte Carlo Bond} (1924) could be noted as precursors to both conceptualism and crypto art alike. Duchamp created \textit{Tzanck Check} to pay his dentist, Daniel Tzanck. It was guaranteed by the (fictive) \textit{Teeth's Loan and Trust Company, Consolidated, 2 Wall Street} for one hundred and fifteen dollars and looked like an ordinary cheque, except for the fact that it was drawn by hand. \textit{Monte Carlo Bond} similarly imitates a financial document, a bond in this case, which was issued by a company established by Duchamp, encouraging investment in his roulette gambling project. The vouchers were signed by his alter ego, ``Rrose S\'{e}lavy, president," and Marcel Duchamp, who is marked as one of S\'{e}lavy's administrators. Thirty bonds were issued but only twelve were sold, all for five hundred francs a piece and buyers were authorized to a yearly dividend of twenty percent. The value of the bonds, just like the value of the cheque, were positioned in relation to the value of the object as an artwork, highlighting the murky definitions of art versus ``non-art," and art versus finance, inquiries that were continued by conceptual artists. 

As art dealer and publisher of conceptual art from the late-1960s through the seventies, Seth Siegelaub built a particular contract architecture for artists whose practice centered on paperwork. Known as \textit{The AC} (\textit{The Artist’s Contract}), it instituted a payment mechanism for artists. Some conceptual artists have used Siegelaub's model to translate the information traditionally present in aesthetic art objects to the contract itself. For instance, conceptual artist Sol LeWitt continues to use contracts in order to provide instructions and sell the rights to execute individual works from his \textit{Wall Drawings} series [1967-2007]. As LeWitt declared in his famous ``Paragraphs on Conceptual Art," from 1967:  \begin{quote}\textit{When an artist uses a conceptual form of art, it means that all of the planning and decisions are made beforehand and the execution is a perfunctory affair. The idea becomes a machine that makes the art.}\end{quote} The artist's job, then, is merely to develop an idea and diagram, to which he attaches a contract sanctioning its production (or reproduction) in visual form. 

LeWitt's use of the \textit{The AC} is indeed comparable to how crypto art works. Both exist first of all  as code, which is a linguistic model or algorithm that is then translated into an image. If, as critic Claire Bishop notes, ``the digital is [..] a garbled recipe of numbers and letters, meaningless to the average viewer," the digital image renders this code (or set of instructions) as a visually perceptible entity \cite{Bishop}. Second, what the blockchain does is make these bits of codes unique and organize them in such a way that they can be traced and tracked. 

Importantly, Siegelaub seemed to have miscalculated the resistance to \textit{The AC’s} adoption, as the majority of the art world rejected the agreement. Its short-lived relevance echoes the general optimism surrounding the revolutionary possibilities of conceptual art and how most of this sentiment was never realized. This exposes how difficult it was (and still is) to successfully institute and formalize artists' rights during financial transactions and art sales. The power that collectors, patrons and galleries have only confirms the degree to which the actor with more wealth ultimately dictates the conditions of exchange. In this way, crypto art seems to solve the objective initially pursued by Siegelaub, by allowing the crypto artist to control the means of distribution. 

If art as a representational system is a helpful model for understanding the (imagined) faith we have in currency as a representation of value -- be it in the form of traditional coins, paper, or electronic fund transfers -- perhaps crypto art could be said to reveal the inner logic of the blockchain, as well the broader financial and art system today. Lippard once wrote that crucial to conceptual art was the examination of information and systems. Today, the blockchain has precipitated an overhaul of the means by which information is transported and represented, while crypto art has become the new (artistic) method of communicating and expressing this structure's dynamic in a visual form.

Nevertheless, in certain crucial respects, crypto art inverses the critique of art’s economic value that was initiated by Duchamp and espoused by practitioners of conceptual art. Conceptual artists sought to get rid of the material object in order to disrupt the art-to-market axis, while crypto artists utilize a similar tactic of dematerialization in order to create something that can be traded more easily. We might conclude, then, that crypto art reveals the logical, if also paradoxical, outcome of the conceptual art strategy. By replacing material commodities with ideas (or code in the form of digital pictures), an ecosystem for the higher velocity of circulation -- that is, a more efficient art market -- has been produced. 

\subsubsection{Further thoughts and provocations}

Crypto art is often positioned as a radical alternative to traditional or conventional art, but when examining its form, we note it has tended to follow the rather academic logic of the two-dimensional, rectangular picture (like a painting or drawing). Since the work is made to be viewed through a screen -- on a computer, phone, or tablet -- much of crypto art work features bright colours and a central, large, bold image. As far as we know, there have been relatively few engagements among crypto art practitioners with the implicit value system of the blockchain. One exception we have identified is \textit{Plantoid},\footnote{\url{http://www.plantoid.org/#history}.} which was developed in 2015 by Primavera De Filippi, the founder of the art collective Okhaos. As Jess Houlgrave has noted, \textit{Plantoid} ``lives" and ``reproduces using Bitcoin micropayments."\cite{Houlgrave} Every Plantoid has its own Bitcoin wallet, and spectators can pay the object in the cryptocurrency, via a QR code, and it will respond by displaying ``colours and sounds." When the plant has collected enough Bitcoin, it ``commissions, through smart contracts another artist or group of artists to produce another plant" on the Ethereum blockchain, funding this through the Bitcoin it has collected. This work is an attempt to atomize patronage and aestheticize the blockchain medium itself. We hope that the modes and boundaries of the blockchain for crypto art production will be pushed even further (a possibility that is mentioned in Section \ref{sec:martin}.). 

Moreover, we note that while the crypto world's emphasis on anonymity can enable radical critiques -- for example, of the art system's emphasis on brand-name artists -- it can also foster a refusal to take responsibility for a specific action or choice. A debate -- and art in general, at its best, is precisely that -- means an artist or critic needs to take, and be accountable for, a position. Without historical and critical reflection, might crypto art even run the risk of dying out as a mere trend?  With this question in mind, we would advocate for the inclusion of an informed critic system within the crypto art world -- that is, a semantic space of dialogue, qualitative judgment, and critique -- perhaps situated on the blockchain itself.

\subsection{The Artists}

\subsubsection{Martin Lukas Ostachowski}\label{sec:martin}

Discovering blockchain technology has been pivotal in my artistic practice. As a technology which is expected to change our lives just like the telephone or the internet has already, I started exploring the blockchain conceptually and as my subject matter using artwork to make this complex technology accessible. 

A few start-ups went practically unnoticed during a period of numerous high-profile Initial Coin Offerings and skyrocketing cryptocurrency values, as they developed some of the first fully-functional use cases for the blockchain focused entirely on living artists: crypto art platforms. These platforms addressed one of digital artists’ biggest obstacles: the perceived inflated value of digital art due to (un)authorized replications. They also made the tokenization for artists easily accessible. Hence, crypto art platforms paved the way for artists to capitalize their digital creations.

\paragraph{A Digital Art Forum}

Crypto art develops an entirely new audience of art enthusiasts. Conceptually linked to the blockchain, crypto art and crypto-related physical art attract the interest of the globally growing crypto community. Numerous conferences showcase or auction off crypto art, regularly include pop up galleries and help promote this new art form. This practice starts cultivating a new form of digital native art collectors, who are familiar with the concept of owning and valuing digital assets. Digital natives appreciate crypto art's appealing character such as its relevance and its accessibility in comparison to traditional art outlets.

Consequently, crypto art expands the limited opportunities to showcase and to sell digital art at events, conferences and through the crypto art platforms. This is both attractive and important, particularly for emerging artists, as they face the traditional art world's bias towards newcomers. Emerging artists have to build up their reputation with smaller exhibitions and events in a lengthy often multi-year process until they start being accepted into larger and more popular events. This tedious process impedes the funding and compensation opportunities for artists at their crucial early career stage. The lengthy practice forces many artists to either supplement their art career with other jobs or to give up. It is easier to realize the significance of early-career funding when one considers that the average artist's income in Canada is, at only \$15,000, almost 60\% lower than the national median \cite{Hill}. Online art galleries and social media brought some relief to this mostly pre-defined process, but focus primarily on physical art. The plethora of artists and artwork and the inevitable competition on an international scale make it more difficult to gain ground. However, when it comes to digital art, the opportunities to showcase and sell were almost negligible before the appearance of crypto art platforms.

Lastly, crypto art has the potential to drastically expand the reach of digital art into the traditional art world. Even almost 50 years after the first digital artworks were recognized as such, the traditional art world still struggles to embrace digital art. Art Basel's global director Marc Spiegler writes in an article that ``For the core of the art world, most digital art seemed overly enamored with its own technology, and often felt conceptually lightweight." \cite{Spiegler} Curator and author Stein van der Ziel criticizes that digital artists are often forced to reduce their digital concepts into shallow -- often static -- physical forms to exhibit in traditional art venues \cite{Stein}. While it remains unclear what causes this lack of appreciation, some of it originates in the general conservatism of the traditional art world, its technical unawareness and especially the impossibility to `commodify' digital assets. Having made transferable ownership possible, digital art can now become a commodity and has the potential to increase the interest of digital art within the traditional art world and even attract individual and institutional art collectors over time.

\paragraph{An Alternative Marketplace}

Aside from the possibility to tokenize digital art, most crypto art platforms are at the same time crypto art market places which connect artists with buyers. For emerging and established artists alike, these platforms simplify the whole sales process tremendously and entirely eliminate many risks and concerns of online sales.

As an example, when an artist starts selling in a traditional online art gallery, typically the physical artwork has to be photographed and edited. After the sale, it has to be packaged and shipped with insurance. The artist has to wait for their artwork to arrive to the buyer and for a variable grace period to pass since most online shops offer unconditional returns. Only after this period expires, the payment is released to the artists, often stretching the whole transaction to a month or beyond. In contrast, on crypto art platforms the artist:
\begin{itemize}
	\item is paid instantly without delays;
	\item benefits from low or no commissions;
	\item benefits from a percentage on all secondary sales (on some platforms);
	\item has full price control and is not subject to promotional discounts unless they are self-initiated;
	\item is exempt from returns;
	\item has not packaging and shipping to consider;
	\item can sell artwork internationally without shipping/customs restrictions;
	\item has the option to remain anonymous;
	\item is exempt from concerns regarding the representation due to colour or texture (what you see is what you get);
	\item is able to gift or exchange artwork freely;
	\item is able to capitalize on already existing physical artwork by digitizing it to an altered, enhanced or even unedited state.
\end{itemize}

Crypto art platforms therefore make the tokenization of digital art simple and accessible. With the few opportunities to sell digital art outside of legally complex licensing agreements and fairly rare digital art commissions, crypto art platforms provide an attractive and new income opportunity. In addition, crypto art platforms are inclusive enablers as they do not focus on the top market segment and encourage emerging artists to create and get compensated. In its early stages, the crypto art market is still evolving and seems to have not yet reached the saturation level of traditional online art galleries.

\paragraph{A multidisciplinary collective}

One of the most remarkable aspects of crypto art is the distinct community built around the crypto art platforms. Most platforms provide public conversation groups in chat applications like Telegram and Discord enabling the exchange and communication between crypto artists, art collectors, enthusiasts and the platform's founders in an unprecedented way. The groups participants are diverse, international and come from a wide variety of backgrounds such as developers, scientists, enthusiasts and artists of all career stages and ages. Moreover, these groups form encouraging communities with artists critiquing each other, collaborating on new artwork or collectors commissioning artists. In these early stages and with existing communities under 200 participants, these groups form multidisciplinary art collectives expanding into real-life projects such as exhibitions within traditional art venues. 

\paragraph{Parenthesis: A Year of Crypto Art}

\begin{figure}[h]
\centering
\includegraphics[scale=0.11]{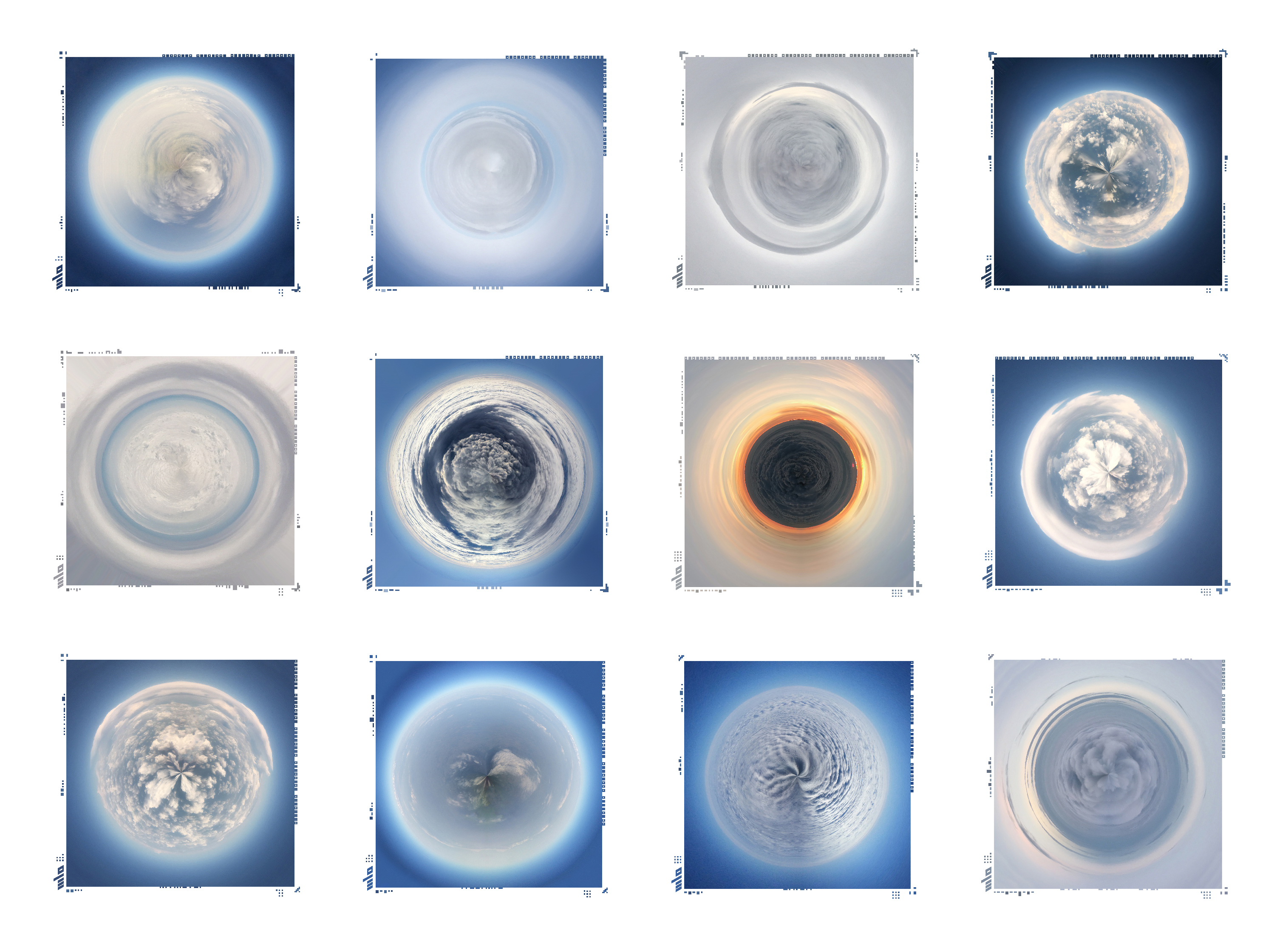}
\caption{Tokenized Cloud Spheres, Martin Lukas Ostachowski, 2018.}\label{fig:martin_spheres}
\end{figure}

After creating crypto art for a year, I noticed the constant intersection of my physical and digital pieces changed my artistic practice and started dissolving the boundaries between the different mediums. An example of this intersection is my installation of twelve photo collages: \textit{Tokenized Cloud Spheres} (Figure \ref{fig:martin_spheres}).

The installation consists of twelve \textit{Cloud Spheres}, which were created from photo series taken on passenger flights across North America. The photos were stitched into panoramas using software and lightly retouched on the edges to allow smooth connections into a sphere. Subsequently, I developed individual borders to transform the photos to visual markers much like a QR code, which can be recognized and overlayed with linked videos through the Artivive Augmented Reality app. I adapted these results initially intended for the crypto art pieces to the limited print editions. 

Aside from using the results of exploring generative art and augmented reality for physical pieces, I also find myself revisiting physical pieces and seeking elements to animate and to create crypto art. Moreover, I often create physical pieces with potential digital enhancements already in mind, making me conscious of the increasing influence crypto art has on my creative process and work. It fuels my enthusiasm to continue promoting crypto art in particular in traditional art venues such as the Czong Institute for Contemporary Art in South Korea, in which a few of my \textit{Tokenized Cloud Spheres} will be exhibited at a photography exhibition later this year.

\paragraph{Conclusion}

Crypto art is a contemporary digital art forum and a relevant addition to the traditional art world. Being able to endlessly replicate digital art and the lack of provable ownership deprived artists from capitalizing their digital creations. The ability to create digital originals has the potential to attract traditional art collectors. Although blockchain technology does not necessarily lead to the recognition of digital art in the art world, it promises a paradigm shift and could be the vehicle to promote digital art within the traditional art world. The public information on art sales, ownership and ownership transfers promote transparency and might influence the mostly unregulated and opaque traditional art market.

The impact crypto art has in regard to early career funding of artists is both remarkable and underrated. Along with its tightly-knit digital art collectives, it seems to drive digital and traditional artists alike to enter the crypto art landscape and to fuel new digital creations. Crypto art has the potential to establish itself as an art movement, not only a temporary trend. Seeing the growth of the blockchain communities undeterred by low cryptocurrency market values is promising.

\paragraph{Open questions}

I would like to point out a few suggestions, possible problems and limitations of crypto art and crypto art platforms from an artist's perspective:

\begin{itemize}
	\item \textit{Centralized processes on decentralized platforms}: One of the core ideas of Blockchain environments is its decentralized nature which democratizes concentrated decision-making powers. This does not entirely apply to crypto art platforms. As an example, almost all platforms still manually approve artists to their platforms. This contradicts the Blockchain philosophy and remains in place due to previous abuse of contributors and the possibility to remain anonymous.
	
	\item \textit{Overtokenization}: The accessibility of crypto art seems to be both a blessing and a curse. As crypto art platforms gain popularity, a few artists started using software to quickly manipulate photos and to tokenize several pieces per day. This will challenge crypto art platforms into better curation to remain relevant market places and not turn away artists and art collectors.
	
	\item \textit{Censorship of controversial art}: There haven’t been notable cases of controversial crypto art which does not exclude them in future. It is questionable how crypto art platforms will treat cases of utterly disturbing or illegal depictions and to which extent they commit themselves to the immutable character of the blockchain. Along with the earlier mentioned overtokenization, this issue might force artists to abandon crypto art platforms to protect their reputation.
	
	\item \textit{Locked Pricing Mechanism}: Unlike lead (fiat) currencies, cryptocurrencies are still subject to significant price fluctuations which can account to 3-digit changes over only a few weeks. Most platforms promote the price in their underlying cryptocurrency. This creates a substantial barrier in the onboarding process of new collectors as well as traditional artworld, especially when the prices are high as it incentivizes existing cryptocurrency holders. In conversations at exhibitions, I predominantly hear that collectors expect to benefit from higher cryptocurrency value and non-crypto art enthusiasts feel alienated with referring to the current spot market price in their fiat currency. It is questionable if artists will revise their pricing during high fluctuation periods or if platforms might consider locking the sales price to the fiat values.
	
	\item \textit{Overarching token standards and crypto art registry}: At the current time of writing, there are several token standards on the Ethereum Blockchain alone, among others ERC-20, ERC-721 and ERC 1155. With the first duplicate cases on different art platforms, the different token standards the platforms use is a challenge. It forces the crypto art community to act as the watchdog for duplicates and abuse of attribution and ownership. As the crypto art movement is growing and more artists start tokenizing art, an overarching standard and registry which the platforms would validate before tokenization could provide a scalable alternative.
	
    \item \textit{Comprehensive expansion of secondary market sales commissions}: While secondary sales often account for a multiple of the initial sales value, the artists usually go away empty-handed. Some platforms try to change that and offer a percentage on all secondary sales to artists. However, this practice is currently limited to sales within the respective platform. The comprehensive extension across all art platforms and token standards could finally reward the actual creator.
    
    \item \textit{Fine art tokens}: Several blockchain start-ups work on the introduction of authenticity and ownership records for fine art. As it adds value to the artwork, these records should naturally be extended for both physical and crypto art to capture even more relevant meta information such as exhibition histories, prizes and notable mentions.
    
    \item \textit{Limiting simultaneous display}: It would be desirable to develop the rare ownership aspect of crypto art even further through the limitation of simultaneous displays of the same artwork. This would facilitate traditional art world use cases such as lending crypto art for art exhibitions and museums.
	
\end{itemize}


\subsubsection{Sergio Scalet (Hackatao)}\label{sec:hackatao}

\begin{quote}
\textit{Cryptoart: \\
The new art history Frankenstein. \\
Baby-monster, Dada-GAN, GIFbit PopFuturist, hyperspeed crypto-evolved deep-decentralized, markedly self-generative.}
\end{quote}

As the Hackatao duo, we had the opportunity to participate in the launch of SuperRare, and be among the first artists who started to tokenize there. We convincingly engaged with this new market and medium of expression. Within the span of a short year, we realized and tokenized 60 artworks (with 2 of them composed of 7 related pieces each), selling 51 of them. The economical gain, while not comparable with that of our traditional artwork, is nevertheless important to keep us involved in crypto art. Yet, we find in new relationships and artistic exchanges the main drives keeping our decentralized experience engaging.

Hackatao starts from a solid and significant experience in the italian market of `tangible' art. We took part in the growing artistic movements of the neo-pop and the Italian Pop Surrealism over the years 2007 to 2012. The concomitant economic crisis unfortunately took its toll on these movements, disbanding most of them. We nevertheless persevered, until our encounter with crypto art in the early 2018. 

\begin{wrapfigure}{l}{6.5cm}
\label{wrap-fig:3}
\includegraphics[width=6.5cm]{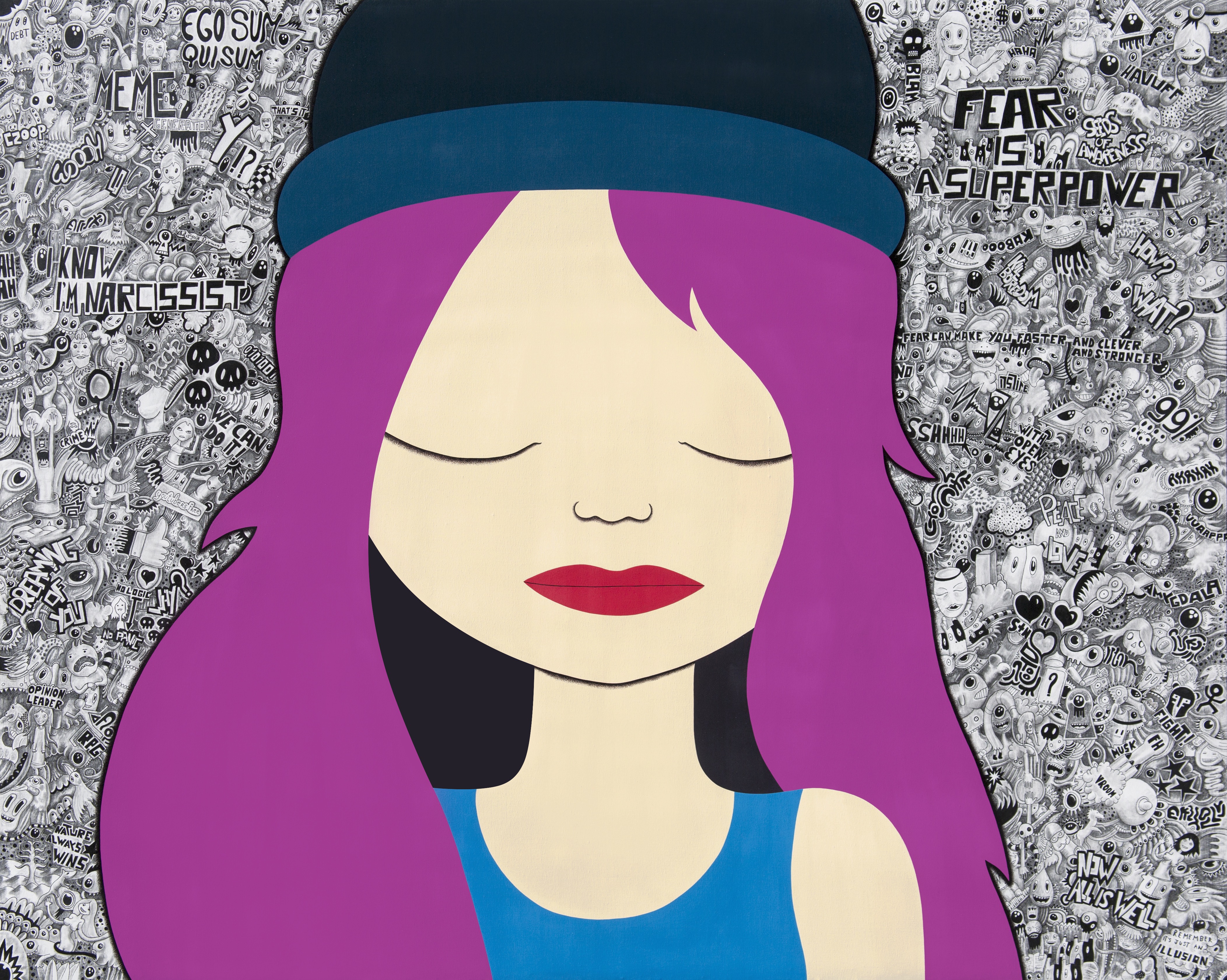}
\end{wrapfigure} 

We like to compare the creative and experimental energies released by crypto art to Dadaism in the 1916s and 20s and, even before, to Duchamp’s `ready-made' paradigm (also see Section \ref{sec:geneaology}). Crypto art gives us a way to freely exercise our skills in a novel combination, blending tangible art with digital media. We can mix pencils, inks and brushes with the Adobe suite to transform our art into the digital. Our research work in crypto art proceeded from our physical artworks which, thanks to their superflat pop backgrounds, can easily be digitally animated. We initially focused on giving a double life to our artworks, by creating digital dopplegangers. Some animated GIFs are linked to their physical counterpart via augmented reality, in so doing preserving their original bond. The very use of augmented reality is meant as a conduit to crypto art for the traditional collectionist, by proposing an involving and stimulating experience pointing them towards digital collecting. It is also worth singling-out our collaboration with the artist HEX0x6C, with whom we created the Hack6C collective. We experimented with manipulating elements taken from our physical art by using algorithms to create novel artworks. These efforts allowed us to explore a creative language with enormous potential in crypto art, generative algorithms, which we were not familiar with.

Our first animated artwork was created back in 2013, out of experimental attempts devoid of a broader goal, and was published on social networks as a GIF. Never could we have imagined that this first artwork, within just a few short years, was going to be tokenized and exchanged as crypto art. In a world used to traditional art, one which is good for hanging on the wall, crypto art represents for us the breaking of two distinct paradigms:
Digital, or crypto artworks, which are infinitely reproducible and as such not of interest to collectors, can become rare thanks to the blockchain, hence being exchanged at a certain value.
The ownership of a digital, intangible artwork, which all can see but only one can own. 

Crypto art further allowed us to explore with means of expression at a much increased pace, when compared with our traditional media of canvas or pottery. Given our style and painstaking attention in both drawing and the use of colour, our paintings take a significant amount of time to complete, sometimes up to months. Crypto artworks, instead, satisfy a primal and immediate need: the urge of an idea translates into reality, marketing and often even sale in a matter of hours. Furthermore, we can fill in the need, shared with many other artists, to stay digital for the full lifecycle of an artwork. Born-digital artworks before crypto art had to be ported to other supports, such as prints, or experienced via ad-hoc installations, losing their original purely digital nature. Nowadays we can experience and trade in artworks with a mobile phone.

\begin{wrapfigure}{l}{6.5cm}
\label{wrap-fig:1}
\includegraphics[width=6.5cm]{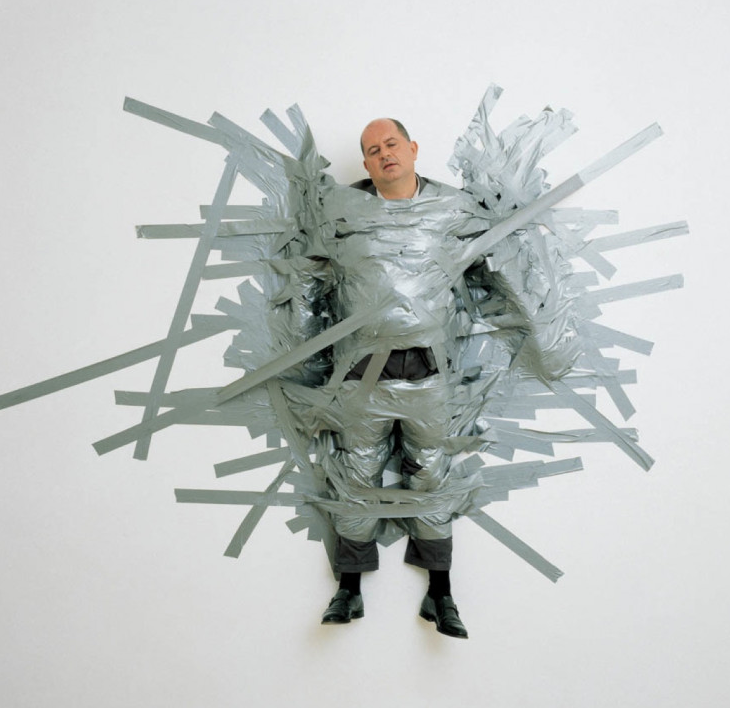}
\end{wrapfigure} 

With traditional art, the creation of an artwork is often a slow process, its exhibition requires time and resources, not to mention a willing gallery. Exposure is further limited by the opening times of the gallery or museum. An artwork can ultimately take years to reach the market, after potentially several changes of hand. Crypto art subverts all this. All artworks are always on display and on the market, both primary and secondary, despite the fact that this latter has not yet developed properly, perhaps as a consequence of hyperinflation or lack of buyers.

The sheer speed of crypto art is to us reminiscent of the words of Tyrell from Blade Runner (1982): ``The light that burns twice as bright burns half as long -- and you have burned so very, very brightly, Roy." We might be in need for a pause, a reflection to bring us on the right track.

After more than a year of involvement in the crypto art world, new questions emerge, whose main thread boils down to: is there room for crypto art in the art system? Instinctively, we say yes. Yes, because crypto art gives voice to new means of expression and languages that are closer to the younger generation of artists. Yes, because it opens the art world to a generation of collectors who do not care for their living room walls, but do engage using mobile devices instead. Cattelan's ``A perfect day" (1999) could be adapted for crypto art by stitching the collector to a mobile screen.

\paragraph{Further thoughts}

What was intriguing during the early days of crypto art, namely the lack of critical selection of artworks from critics and gallerists, is now becoming a necessity to avoid rampant hyperinflation of crypto artworks. The consequences of hyperinflation could manifest in a reduction in the average value of artworks, loss of investors and subsequent move of artists to other, more stable markets. It is becoming a necessity to find means to keep the hyperinflation of artworks at bay. An option could be to introduce a limit to the number of monthly tokenized artworks for every artist, or request them a fee for each piece published in the gallery, a sort of proof-of-quality-work, compelling a mechanism of self-selection on the part of the artists towards higher quality.

Shall we leave this choice in the hands of the market? Have crypto art collectors a need for critical guidance in their choices? We are living in a primordial soup: we need selection and evolution, towards higher order and equilibrium in a system with high creative potential. A possibility is for specialized galleries to emerge, e.g. relying on Pixura\footnote{\url{https://pixura.io}.}, focusing on crypto art AI, generative crypto art, crypto art Meme, crypto art VR, financial and digital crypto art.

The traditional art system, sooner or later ends up assimilating the artistic movements that want to break free from it, much like the biblical Moloch or Chronos devouring his children (Goya's ``Saturn devouring his son", 1819-23). Mythology teaches us that the old gods are superseded by new ones.

It might be that crypto art, for its own good, will have to be assimilated. It has all the feats of the new forms of collecting proper of millennial and Z generations. It is up to us, artists, developers, collectors, gallerists to make crypto art into an experience valuing art and artists first and foremost.

Nowadays, when browsing a digital gallery such as SuperRare or KnownOrigin, we will immediately notice the constant flux of new artworks, and while it might be artistically interesting to experience all this buzzling activity, it is all the same unsettling for the enthusiast or the collector who won’t be able to fully appreciate every artwork. It feels at time like a frantic Instagram experience. If we enter into a renown art gallery, we will notice the presence of few, selected artworks being exposed. This process gives importance and pace to every artwork and its experience by the public. Maybe, we need to find a sweet spot between speed and slowness, a new equilibrium in order not to lose sight of the very meaning of art.\footnote{An example of cross-platform curated experience is First Edition: \url{https://firstedition.art}. \textit{``How do I get a work on First Edition? Post your work on Editional (\url{https://editional.com}) and get it noticed. First Edition will develop more formal acceptance criteria over time, but for now the best criteria is ``exceptional use of the digital medium". We think digital art nonfungibles will be valued, in part, as Schelling points (\url{https://en.wikipedia.org/wiki/Focal_point_(game_theory)}); this suggests a strategy: post the assets that you believe others will pay more for, relative to other such assets."}}

\subsection{The gallerists}

\subsubsection{SuperRare (Jonathan Perkins)}

SuperRare is less of a gallery and more of a live organism. The founders envisioned it from the start to be a social marketplace. Open, dynamic, and expanding in scope, it is a place where creators can fully harness the power of crypto art, and where participants across the globe form a social layer for transacting, valuing, appreciating, and furthering the field of digital art.

The core pieces of the platform are authentication, issuance, exchange, and provenance, all layered on top of a social network. 

For creators, SuperRare is a place to tokenize and host their works in a combined primary and secondary marketplace. All bids, sales, and newly minted works appear in SuperRare users' activity feed, akin to a social platform like Twitter or Instagram. Thanks to the programmable nature of smart contracts, all secondary sales on SuperRare have a built-in percentage that goes back to the creator. This means that when a work repeatedly changes hands and sells many times over the course of a decade, a stream of revenue continues to flow to the artist–something that was simply not possible before.

For collectors, SuperRare is a place to collect, trade, view and manage their entire collection. The beta version of the platform included only works by SuperRare creators, but an increasing number of additional crypto art projects and tokens is being made available. The open nature of tokenized digital artworks makes it natural for them to flow into a multi-asset secondary market, and the vision is to make SuperRare the collector's home for their whole crypto art collection.

What was the underlying motivation for building SuperRare? I see this decade as a unique opportunity that comes perhaps only once in one's lifetime, when a technological advancement enables entirely new capabilities in a field such as digital media and the arts. Crypto art presents a paradigm shift in digital ownership, enabling new revenue streams for creators, improving transparency and market liquidity for collectors, and indeed unlocking entirely new market possibilities. While one can never foretell exactly how a powerful technological shift will play out over the course of a decade, we were compelled to help steer the crypto art spaceship in a direction that will bring more money into the pockets of artists, and bring more creativity into everyone's lives.

One of the main challenges in building a crypto art marketplace is context. How does one evaluate an individual work? How does an individual, or a broader market, see where a work fits into the existing canon of an artist? Being natively digital and internet-based, crypto art needs digital, internet-based solutions to these questions. Our approach has been not to curate or dictate what work we think is valid or valuable, but to create a transparent social environment that makes it easy for market participants to come to their own conclusions. For example, we favor making data available about an artist and their works over providing editorial commentary.

Another challenge is that crypto art is a new idea in an ancient field. Crypto art ownership, while intuitive and natural to the digital generation, is at least partly conceptual ownership and presents a challenge to many people. This is particularly true if a person has a preconceived idea that crypto art somehow encrypts or otherwise hides the art from non-owners, which at least currently it does not. I think this challenge will diminish over time as the space matures, but all of us acting in the industry have a responsibility to spearhead the conversation, agree upon standards and concepts, and clearly articulate them to newcomers.

What is the biggest opportunity to have an impact and accelerate the field of crypto art? I believe it lies in bringing the collector side of the market to maturity. While the artist side of the market is indeed nascent and still growing, there is repeated, proven demand from creators, because of the obvious and immediate incentives and income opportunities. However I am convinced that the market benefits of crypto art are equally powerful for collectors: transparency, provenance, liquidity, social signaling, and online collection management. It is this side of the market that is slightly more nuanced and needs further development, and is thus a primary focus for SuperRare. When both sides of this market are developed and balanced, we will see crypto art gain adoption and flourish into a rich, global phenomenon the likes of which we can only imagine.

For those interested in \textit{SuperRare market data}, we have a beta version of a public GraphQL API available at \url{https://api.pixura.io/graphiql}. We are enthusiastic about making data and insights increasingly available to the public; while we are still working improving the API and documentation, those interested can get in touch with us for specific questions or requests at \url{hello@pixura.io}.

\subsubsection{KnownOrigin (James Morgan)}\label{sec:knownorigin}

KnownOrigin started with an idea in the pub at the end of 2017/start of 2018, by April 2018 we had launched a very early alpha site with several artists and several pieces and kicked it off with a small pop-up gallery in Manchester, UK. Since its inception we have changed the site drastically, introducing new software and services in order to provide a much better user experience for those involved as well as opening up the platform to a self-service model and increasing the number of controls an artist has.
 
When we created KnownOrigin, we had a few basic ideas and assumptions which we wanted to test out. These included an assumption that the digital arts market is a real thing and artists and collectors understand what it means to create and own digitally scarce limited-edition artworks. Another assumption was that KnownOrigin could provide a platform for unknown artists to monetise their work, empowering them in a way which they currently do not have. We also wanted to prove out a use case for using non-fungible tokens, NFTs, as a protocol and mechanism for tokenisation of artwork. Along with these ideas, we wanted to test a more purist approach to the traditional Web 2.0 era. I mean by this that we do not want capture email address, we do not force a collector to sign-up since they only require a web3 wallet and all our source code, both frontend and backend, is 100\% open source as are all our smart contracts.
 
Today KnownOrigin boasts a large array of features such as self-service tokenisation, fiat purchases via credit cards, bidding and offer systems, artist pricing controls, artist gifting controls, limited edition artworks, artwork likes as well as features which allow artists and collectors to easily view sales and purchases. We still have a long way to go and I think we are still in the early stages of the blockchain enabled digital arts movement, so there is much more work to do. Next on our list is to streamline the collecting experience, proving a more seamless approach to purchasing and owning tokenised artworks.
 
\paragraph{On open challenges and opportunities}
 
Within a few months from launching we realised that demand was so high for new artists and artworks we could not scale the manual process of onboarding and listing artworks for everyone. We then shifted focus to the self-service model which is in place today. This means that artists are given more control and creativity by allowing them to upload and tokenise work without our involvement, facilitated by smart contracts and IPFS. Other challenges exist in scaling the site from tens of users to thousands of monthly users whilst still providing a solid experience. We also have seen an appetite for free art work which is great for the community, but as a long-term business not something we want to embrace too strongly. Other technical challenges exist such as image compression and site speed. Servicing uncompressed ultra-high-resolution images of 25mb might entail that site visitors do not have a good enough experience, we thus introduced site image compression services with a slight impact on image quality. We have faced challenges existing in all blockchain-enabled application, such as how to onboard users in interacting with public blockchains and their nuances. We plan to tackle some of these open issues in the near future, by leveraging existing work done by teams around the world who have been focusing on UX and blockchain already.
 
We believe that there still exist silos in the space and we have started to explore newer concepts of how a decentralised autonomous organisation may be leveraged to provide open reputation and registry systems which all platforms can use.\footnote{\url{https://medium.com/knownorigin/arts-focused-daos-7999b558d232}.} We believe that eventually the concept of verified and authenticated artworks will be provided in a more decentralised model so collectors do not have to trust the platform but they can trust other means which are based on a claim, stake or reputation from the original creator and their collectors. Eventually, we would like to see a community-driven approach to tackling common problems in the crypto art system.

\paragraph{Technical challenges in applying a price model}

\begin{enumerate}
\item Owning digital artworks and creation of artworks is as easy as ever but still there exists a problem of on boarding people into the world of cryptocurrencies and decentralisation. Solving this takes time, requires education and advances in technical know how and user confidence. KnownOrigin and SuperRare are making waves in this process of onboarding new artists and collectors but the market in general is still small.
\item Although counterfeiting is trivial to detect when checking a known token for its authenticity, it becomes harder at a macro level. Enforcing this also becomes problematic in a decentralised and trustless world due to the vastness and lack of control after being identified. In the future there could exist decentralised autonomous organisations (DAOs), which can provide a trustless and open solutions where users can easily detect forgeries across the crypto art world.
\item ERC-721 standards facilitate the creation of crypto art but do not enforce the rules beyond token schematics. Concepts such as secondary sales currently require marketplaces to apply this rule, when operating within the interoperable ERC-721 ecosystem: this is not always guaranteed. The fact that it is possible to facilitate this process is still a great achievement and, in the future, we may see better solutions for secondary market sales. 
\item Crypto art and tokenization allow provenance to be carried along artworks, however deciphering it is not trivial for the average user. Advancements in both retrieving and surfacing these data are needed. Even when data are easily surfaced, other obstacles exist such as understanding the flow and internal dynamics of the ERC-721 token itself. In order to do this without significant effort, the smart contract source code ought to be available and verified, something that cannot be enforced.
\end{enumerate}

Technology has given us solutions to known problems which exist in the space but inevitably bring up new ones. As with all new technological changes, artists, users, collectors, galleries and exhibitors must adapt and learn how to operate. Along with the digital world, we are witnessing the combination of crypto art with physical art, which will open up new challenges to all parties involved in the process. In the end, the solutions found will be superior to the existing status quo. \textit{KnownOrigin data are available upon request.}

\subsection{The collectors}

The Internet is increasingly ubiquitous and our lives are turning overwhelmingly digital. Up to 2008, there was no way for anyone to truly ``own" anything on the Internet. For those who grew up playing Everquest and World of Warcraft, the concept of ``ownership" meant that when you logged back into the game, you would still own a certain item in your inventory. Unless the server was down, or completely shutdown by lack of players. Or unless the company went bankrupt or decided to remove that item from gameplay. The creator of Ethereum, Vitalik Buterin, famously wrote:

\begin{quote}
\textit{I happily played World of Warcraft during 2007-2010, but one day Blizzard removed the damage component from my beloved warlock's Siphon Life spell. I cried myself to sleep, and on that day I realized what horrors centralized services can bring. I soon decided to quit.}
\end{quote}

Smart contract platforms allow game developers to migrate the definition of ownership from their game server to a decentralized trustless network such as Ethereum. When the game starts, the player's inventory is not defined by what the game company SQL database holds, but by what items the player owns on the Ethereum blockchain. From the gamer's perspective, it's a huge win because they will \textbf{forever} keep the items they earn through hours of gameplay and they can potentially reuse them in other games in the series. There are even instances of games collaborating with one another, and allowing their user base to use the items from one game in the other.\footnote{See the collaboration between the collectible game CryptoKitties  and the trading card game Gods Unchained, \url{https://www.cryptokitties.co/gods-unchained}.} This is unheard of in the traditional gaming market (both from a business and technical point of view). From the gaming company perspective, the obvious downside is that they can't extract rent from direct transfers between players anymore and that they lose monopoly over the secondary market. But lowering the friction to purchase items, the cross-pollination between games and the increase gamers sense of ownership could potentially lead to an increase revenue overall. We might eventually get to a point where a large portion of gamers demand their items exist forever in the blockchain under their own control. A similar situation is happening with collectibles such as sports trading cards.\footnote{See \url{mlbcryptobaseball.com}.}

We are virtual reality (VR) scene developers for a VR world where all ownership is tracked via blockchain: Decentraland. This includes the land itself; a total of 43,689 LAND were sold to ``digital landlords" via smart-contract enabled auctions. Each LAND has a unique x and y coordinates (from -150 to 150), and whoever owns a specific LAND controls what 3D models are displayed on it. It uses the same standard (ERC-721) as other major NFT providers use (SuperRare and KnownOrigin included), which means it works out of the box with many decentralized exchanges and wallets. It is the first time that people can truly ``own" virtual land; they control what's on it, they can resell it, donate it, destroy it, without anyone being able to do anything to stop them. This, combined with a rent market through artificial scarcity, provides the right incentives for landlords, developers and entrepreneurs to build interesting experiences in that world. Our business model is simple: we will build amazing free-to-use VR experiences on a majority of our land (say 80\%), driving traffic to that area. The remaining portion (say 20\%) will be available for commercial use such as renting or running a franchised game/experience. We believe neighborhoods will develop in virtuous cycles, better content driving increasingly better content. We want to contribute positively to that cycle and reap the benefits through the rent market.

Art is an important element of building interesting VR experiences. We are currently building an Art \& History museum to tell the history of Ethereum through notable blocks. Art here is used to complement a textual description of a given historical event (say a hard fork) and make the overall experience more compelling. In future projects, we will use art to enhance certain experiences (for example, waiting in an elevator or on the bus). While it would be possible to find free-to-use images on the internet to populate the scenes, we will be using crypto art because there is a strong synergy between us, the scene developers, and the artists. We will display their artwork (free of charge, in non-commercial setting) and drive content to their marketplace and personal page. In turn, these new eyeballs along with a liquid market will help properly assess the quality of the artists and of their work. If our earlier assessments are right (and we have identified good quality artwork), the price of our NFT should increase. Note that even if our assumptions are wrong (crypto art never takes off, or takes off and dies shortly after, or we picked the wrong artists and artwork), we could still recoup our upfront investment via the rent market on Decentraland (and vice-versa with regards to our assumptions on Decentraland).

\subsubsection{Why we prefer crypto art over physical art}

\begin{itemize}
\item We appreciate the true digital ownership aspect. It means anyone can carry their entire collection with them at any time, it means they can gift it to anyone across the planet. Custody becomes easy.
\item It's never been easier to own and display animated artwork. Owning a painting and displaying it on your wall in the physical world is easy, doing so with a 48 frames animated artwork not so much. Both can be done within Virtual Reality worlds (although resolution and file size limitations can be a hassle).
\item Both the primary and secondary markets are fully observable, transparent, never stop, and do not require the trust of any-party.
\item Crypto art, as all ERC-721 tokens, are programmable. For example, they can be atomically swapped between two parties or they can be used as collateral on loans. 
\item Anyone can bid on any of our crypto art at any point in time. No need to place items for sale.
\item Artists can get properly compensated on secondary sales in a provable, programmatic way.
\item Counterfeit is impossible by virtue of the smart contract platform (although proving authenticity remains a challenge).
\end{itemize}

As collectors, we have been establishing a pricing model. It is work-in-progress, but some aspects include:

\begin{enumerate}
\item Is the creator even real, or simply a bot that steals content from other platforms? For this we use their social network references and web presence in general, as well as the diversity of what they have published so far. This is where the marketplace itself can provide value and do these checks for us (for example ``verified creator" badge). We echo comments from Section \ref{sec:knownorigin} regarding the possible use of DAOs and Token-Curated Registries to serve this purpose in a more decentralized way. 
\item Is the artwork original, and has it never been tokenized on another platform? There's no easy solution for this right now. There's a need for a reputation system that's built outside of each individual marketplaces and that covers them all. 
\item Assuming the creators are actual humans, are they known artists? How can we tell if an artist is being impersonated? In some cases, we have reached out to artists by email to confirm that their identity and digital artwork were not being misused.  
\item The quality of the artwork. For static images, similar qualities as physical equivalent can be used. For animated artwork, there are a few additional elements (nature of each frame, perfect-loop, frames annotated, transitions between frames, timing of elements in the scene).
\item Is it a good fit for a Virtual Reality gallery? For animated artwork, no strong flashes, proper rythm, proper colors.
\item Social metrics (views and likes). Easily gameable so not a strong indicator.
\item Total supply for that artist. We echo concerns from Section \ref{sec:hackatao} that hyperinflation is an issue and seriously dilutes the collector experience. 
\item The starting price, time, and current bidding history. 
\item Current crypto art market condition. What's the starting bid these days? How many bidders on average? 
\item Current Ether market condition when bids are denominated in ETH. This would not be necessary if the marketplaces would allow for stablecoin-denominated bids and sales (only OpenSea seems to do so as of May 2019). Stablecoins are ERC-20 tokens designed to track the value of a physical asset. A famous example is the DAI \cite{MakerDAO} by the MakerDAO team which tracks the US dollar (1 DAI = 1 USD). 
\item Historical sales data for an artist, especially on the secondary market.
\item Quality of the NFT provider/marketplace. As an example, while we would still keep ownership of our NFT if the SuperRare.co marketplace would shutdown, it would also mean that we would need to rely on secondary marketplaces (such as OpenSea) for both viewing and trading our collections. 
\end{enumerate}

On top of the artwork pricing model, we are continuously assessing the systemic risks to our crypto art investment. In particular, we are watching the NFT adoption trends (across all use cases: gaming, collectibles, artwork, virtual estate), and the development of the underlying blockchain, Ethereum. Many of the signals above need to be manually extracted by us. Having access to a crypto art analytics platform would allow us to better price our bids and manage our investment.

\section{Coda}


We individuate the following areas for ongoing and future work emerging from our decentralized position paper on crypto art: understanding the artistic movement; values; engagement and community; economics; blockchain technology and applications; analytics.

\paragraph{Understanding the crypto-artistic movement} Art scholars' T'ai Smith and Blake Finucane provide much needed historical and critical depth to crypto art, uncovering that while the mix of blockchain technology and digital art might be new, there are antecedents to consider and be informed by. In this respect, crypto art offers to scholars a singularly compelling entry point to study the encounter of art, technology and socio-economic systems in the digital space.

\paragraph{Crypto art values} The crypto art phenomenon is intimately linked to the values that the blockchain technology has come to represent. Crypto art is, in the words of SuperRare gallerist Jonathan Perkins, ``a new idea in an ancient field". Decentralization, democratization and individual control are themes emerging from the viewpoints of both artists, gallerists and collectors. For the artists, crypto art represents in this sense a way to get and keep control of their artworks, and reap related benefits. Sometimes these values are not fully materialized yet, such for example in the case -- discussed by KnownOrigin's James Morgan and artist Martin Lukas Ostachowski -- of the gallery-centralized verification and authentication of artists and artworks.

\paragraph{Engagement and community} Crypto art is seen as an artistic phenomenon most appealing to a broader and younger cohort of potential artists and collectors, including but spanning beyond the crypto community. This point is made explicitly by both artists and gallerists, who have an interest and appreciation for their community of peers and customers. The artists Martin Lukas Ostachowski and Sergio Scalet from Hackatao both describe and exemplify how crypto art allows them to move across the physical and the digital spaces, with a speed and freedom of experimentation previously unknown. The space for exploration is still quite wide if we consider that, until now, crypto artists have been focusing on the relatively traditional format of rectangular images and GIFs, as T'ai Smith and Blake Finucane aptly note. While for the artists crypto art makes for an opportunity to engage with new peers in the digital art space, as well as to attract new buyers, for gallerists the engagement with the community goes through the task of ever perfecting their platforms in terms of services and ease of onboarding. For Jonathan Perkins, the biggest opportunity in crypto art is to bring the collector side of the market to maturity, by designing marketplaces to maximize the benefits of transparency, provenance, liquidity, social signaling, and online collection management. Crypto art might be, in this respect, just the beginning of a whole new way to create, exchange and experience art in the digital space, of which the collector Sebasti\'{a}n Hern\'{a}ndez gives a compelling example discussing art exhibition in the VR world Decentraland.

\paragraph{Economics of crypto art} The economic aspects of crypto art are crucial to the viewpoints of almost all contributors. For the artists, crypto art offers a way to directly market their own artworks without mediation and at unprecedented speed: a crucial possibility especially for new artists, as Martin Lukas Ostachowski underlines. Crypto art's immediate economic incentives, and their sometimes unintended consequences, are discussed and historically contextualized by the art scholars T'ai Smith and Blake Finucane. In a curious twist of history, crypto art has finally `materialized' Duchamp's and the conceptual artists' attempt to get rid of material objects in order to expose and disrupt the traditional art market. In doing so, it has created a much more rapid market for digital artifacts, whose velocity makes it akin to financial trading in some respects, according to the data scientists. Perhaps as a consequence of the high speed and the relative lack of control on artwork publishing, a significant current challenge for crypto artists is artwork hyperinflation (Sergio Scalet) or overtokenization (Martin Lukas Ostachowski). The sheer amount of new artworks does not allow for users and buyers to experience, digest and eventually buy artworks before a flood of new creations arrives. Hyperinflation has not only the effect of depressing prices, but also of preventing a secondary market to emerge. A possible step forward could be to propose curated experiences, either directly by third parties and processes or using Decentralized Autonomous Organizations (DAOs) and Token-Curated Registries (TCRs).

\paragraph{Blockchain technology and applications} While Crypto Art is fully engaging with crypto technology, there is a shared agreement that the movement has not fully tapped into the potential of blockchain technology and value system. A relevant point has been made regarding the use of crypto currencies whose volatility makes the market, and collectors' behaviour in particular, more difficult to analyze and predict. Locked pricing mechanisms and the adoption of stablecoins might offer a way forward in this respect, as discussed by Martin Lukas Ostachowski and Sebasti\'{a}n Hern\'{a}ndez. Another striking example is the lack of interoperability among galleries, which use different standards and identification of artists and artworks. This has the direct effect of currently limiting the secondary market to within-gallery exchanges. Another, related problem is the system-wide verification of counterfeiting: a solution across the crypto art world might come from DAOs and TCRs in this case as well. System-wide interoperability would also allow for extra-system (re-)use of crypto artworks, as well as for the creation of third-party services such as a crypto art analytics platform proposed by Sebasti\'{a}n Hern\'{a}ndez.

\paragraph{Crypto art analytics} A crypto art analytics platform would require first and foremost the collection and alignment of data from different galleries or, more conveniently, their adoption of a shared standard to expose them programmatically. It would further necessitate the development of crypto art metrics informative to the different users of the platform, such as artists, collectors, gallerists and even by-standers. This is a direction for future work proposed by data scientists Massimo Franceschet and Giovanni Colavizza. Furthermore, the open availability of crypto art data, or ``context", will ultimately allow data scientists to model the transaction system and predict the future success of individual artworks and artists. Crypto art offers the full availability of artwork data (images and metadata), transaction data (bids and sales) and social data (likes and views), and thus will allow to study the mechanics of success in art and creative industries, perhaps at unprecedented detail.

\bibliographystyle{plain}

\begin{thebibliography}{10}

\bibitem{ipfs}
{IPFS}.
\newblock \url{https://ipfs.io}.
\newblock Accessed: 2019-02-08.

\bibitem{knownorigin}
{Known Origin}.
\newblock \url{https://knownorigin.io}.
\newblock Accessed: 2019-02-08.

\bibitem{superrare}
{SuperRare}.
\newblock \url{https://superrare.co}.
\newblock Accessed: 2019-02-08.

\bibitem{Bishop}
C.~Bishop.
\newblock Digital divide: Contemporary art and new media.
\newblock {\em Artforum}, 51(1), 2012.

\bibitem{Buchloh}
B.~Buchloh.
\newblock Conceptual art 1962-1969: From the aesthetic of administration to the
  critique of institutions.
\newblock {\em October}, 55:105.

\bibitem{FSRRB18}
S.~P. Fraiberger, R.~Sinatra, M.~Resch, C.~Riedl, and A.-L. Barab{\'a}si.
\newblock Quantifying reputation and success in art.
\newblock {\em Science}, 362(6416):825--829, 2018.

\bibitem{HS91}
S.~Haber and W.~S. Stornetta.
\newblock How to time-stamp a digital document.
\newblock {\em Journal of Cryptology}, 3(2):99--111, 1991.

\bibitem{Snail}
HEX0x6C.
\newblock {O Snail}.
\newblock {\em SuperRare}, 2018.
\newblock Available on \url{https://superrare.co/artwork/o-snail-1245}.

\bibitem{Houlgrave}
J.~Houlgrave.
\newblock {\#BlockchainArt}.
\newblock
  \url{https://medium.com/codexprotocol/https-medium-com-jessyblock-blockchainart-de4981678412},
  2017.
\newblock Accessed: 2019-06-01.

\bibitem{Hill}
Hill Strategies~Research Inc.
\newblock A statistical profile of artists in canada in 2016 (with summary
  information about cultural workers).
\newblock
  \url{https://hillstrategies.com/resource/statistical-profile-of-artists-in-canada-in-2016},
  2019.
\newblock Accessed: 2019-03-19.

\bibitem{Kosuth}
J.~Kosuth.
\newblock Art after philosophy.
\newblock \url{http://www.ubu.com/papers/kosuth_philosophy.html}, 1969.
\newblock Accessed: 2019-05-04.

\bibitem{Krauss}
R.~Krauss.
\newblock {\em Originality of the Avant-Garde and other modernist myths}.
\newblock MIT Press, Cambridge (MA), 1985.

\bibitem{3V}
D.~Laney.
\newblock 3d data management: Controlling data volume, velocity, and variety.
\newblock
  \url{https://blogs.gartner.com/doug-laney/files/2012/01/ad949-3D-Data-Management-Controlling-Data-Volume-Velocity-and-Variety.pdf},
  2001.
\newblock Accessed: 2019-05-04.

\bibitem{Lippard2}
L.~Lippard.
\newblock {\em Escape attempt. Six years: The dematerialization of the art
  object from 1966 to 1972}.
\newblock Praeger, New York, 1973.

\bibitem{Lippard}
L.~Lippard and J.~Chandler.
\newblock The dematerialization of art.
\newblock {\em Art international}, 11, 1968.

\bibitem{LWSGSW18}
L.~Liu, Y.~Wang, R.~Sinatra, C.~L. Giles, C.~Song, and D.~Wang.
\newblock Hot streaks in artistic, cultural, and scientific careers.
\newblock {\em Nature}, 559(7714):396--399, 2018.

\bibitem{MI18}
B.~Mitali and P.~L. Ingram.
\newblock Fame as an illusion of creativity: Evidence from the pioneers of
  abstract art.
\newblock Technical report, HEC Paris Research Paper No. SPE-2018-1305, 2018.
\newblock Available from \url{http://dx.doi.org/10.2139/ssrn.3258318}.

\bibitem{Nakamoto08}
S.~Nakamoto.
\newblock Bitcoin: A peer-to-peer electronic cash system.
\newblock 2008.
\newblock Available from \url{https://bitcoin.org/bitcoin.pdf}.

\bibitem{Roeder}
O.~Roeder.
\newblock The blockchain is just another way to make art all about money.
\newblock
  \url{https://fivethirtyeight.com/features/blockchain-is-just-another-way-make-art-all-about-money},
  2018.
\newblock Accessed: 2019-05-04.

\bibitem{SPR18}
H.~Y.~D. Sigaki, M.~Perc, and H.~V. Riuchloeiro.
\newblock History of art paintings through the lens of entropy and complexity.
\newblock {\em Proceedings of the National Academy of Sciences},
  115(37):E8585--E8594, 2018.

\bibitem{Spiegler}
M.~Spiegler.
\newblock From selfies to virtual reality, how technology is changing the art
  world.
\newblock
  \url{https://www.cnn.com/style/article/marc-spiegler-cnn-style-guest-editor/index.html}.
\newblock Accessed: 2019-03-19.

\bibitem{MakerDAO}
The~Maker Team.
\newblock The {DAI} stablecoin system.
\newblock \url{https://makerdao.com/whitepaper/Dai-Whitepaper-Dec17-en.pdf},
  2017.
\newblock Accessed: 2019-05-27.

\bibitem{Stein}
S.~van~der Ziel.
\newblock The frustrating way the art world deals with digital art.
\newblock
  \url{https://medium.com/datadriveninvestor/the-frustrating-way-the-art-world-deals-with-digital-art-9450f101485}.
\newblock Accessed: 2019-03-19.

\bibitem{NIST18}
D.~Yaga, P.~Mell, N.~Roby, and K.~Scarfone.
\newblock Blockchain technology overview.
\newblock Technical report, National Institute of Standards and Technology,
  2018.
\newblock Available from \url{https://doi.org/10.6028/NIST.IR.8202}.

\end{thebibliography}

\end{document}